\documentclass[%
 preprint, 
superscriptaddress,
preprintnumbers,
 amsmath,amssymb,
 aps, physrev,
prb,
]{revtex4-2}

\usepackage{graphicx}
\usepackage{dcolumn}
\usepackage{afterpage}
\usepackage{bm}
\usepackage{ulem}
\usepackage{hyperref}

\begin{document}


\title{\textbf{Measuring superconducting arcs by ARPES}}%

\author{A. Kuibarov}
 \email{a.kuibarov@ifw-dresden.de}
\affiliation{Leibniz Institute for Solid State and Materials Research, IFW Dresden, 01069 Dresden, Germany}

\author{S. Changdar}
\affiliation{Leibniz Institute for Solid State and Materials Research, IFW Dresden, 01069 Dresden, Germany}

\author{A. Fedorov}
\affiliation{Leibniz Institute for Solid State and Materials Research, IFW Dresden, 01069 Dresden, Germany}
\affiliation{Helmholtz-Zentrum Berlin für Materialien und Energie, BESSY II, 12489 Berlin, Germany}
\affiliation{Joint Laboratory ``Functional Quantum Materials`` at BESS II, 12489 Berlin, Germany}

\author{R. Lou}
\affiliation{Leibniz Institute for Solid State and Materials Research, IFW Dresden, 01069 Dresden, Germany}
\affiliation{Helmholtz-Zentrum Berlin für Materialien und Energie, BESSY II, 12489 Berlin, Germany}
\affiliation{Joint Laboratory ``Functional Quantum Materials`` at BESS II, 12489 Berlin, Germany}

\author{O. Suvorov}
\affiliation{Leibniz Institute for Solid State and Materials Research, IFW Dresden, 01069 Dresden, Germany}

\author{V. Misheneva}
\affiliation{Helmholtz-Zentrum Berlin für Materialien und Energie, BESSY II, 12489 Berlin, Germany}
\affiliation{Joint Laboratory ``Functional Quantum Materials`` at BESS II, 12489 Berlin, Germany}

\author{L. Harnagea}
\affiliation{Leibniz Institute for Solid State and Materials Research, IFW Dresden, 01069 Dresden, Germany}

\author{I. Kovalchuk}
\affiliation{Kyiv Academic University, 03142 Kyiv, Ukraine}

\author{S. Wurmehl}
\affiliation{Leibniz Institute for Solid State and Materials Research, IFW Dresden, 01069 Dresden, Germany}


\author{B. Büchner}
\affiliation{Leibniz Institute for Solid State and Materials Research, IFW Dresden, 01069 Dresden, Germany}
\affiliation{Institute for Solid State and Materials Physics, TU Dresden, 01062, Dresden, Germany}

\author{S. Borisenko}
\affiliation{Leibniz Institute for Solid State and Materials Research, IFW Dresden, 01069 Dresden, Germany}

\date{\today}

\begin{abstract}
Angle-resolved photoemission spectroscopy is the leading tool for studying the symmetry and structure of the order parameter in superconductors. The recent improvement of the technique made it possible to detect the superconducting energy gap at the surface of topological t-PtBi$_2$ via observation of the record-breaking narrow line shapes. The promising new physics uncovered requires further investigation of the spectral and gap functions of t-PtBi$_2$, but the challenging experimental conditions severely limit the application of conventional ARPES setups. In this work, we use synchrotron-based measurements and show that the gap at the surface Fermi arc in t-PtBi$_2$ can be detected even with more relaxed experimental conditions than in our previous laser-based studies. At the same time, using simple model of ARPES spectra, we identify the minimum requirements to detect the gap and consider cases where the gap cannot be resolved.

\end{abstract}

\maketitle


\section{\label{sec:level1} Introduction}

Research aimed at discovering materials that simultaneously host topological and superconducting features has recently gained enormous attention in solid-state physics due to their potential to host Majorana fermions, making them applicable to fault-tolerant quantum computing \cite{Kitaev2001}. The type-I Weyl semimetal PtBi$_2$ is one such promising candidate. Unlike many previously proposed systems, where topological superconductivity must be engineered via heterostructures, t-PtBi$_2$ has the potential to be an intrinsic topological superconductor — a system in which these exotic properties can be realized within a single material.

Our previous ARPES report \cite{OurNature}, as well as STM findings \cite{SchimmelSTM, SchimmelQPI}, have shown that in t-PtBi$_2$, a superconducting gap opens exclusively on the topological surface states — the Fermi arcs of the Weyl semimetal with a critical temperature $T_c = 8 – 14$ K, while the bulk bands remain in the normal state down to $0.6 – 1.1$ K \cite{VeyratTransport, Shipunov2020}. This coexistence of a superconducting surface and a non-superconducting bulk presents an experimental challenge, as only surface sensitive techniques, such as those mentioned above, can detect the superconductivity. ARPES is particularly powerful in this context, as it allows one to directly observe the electronic dispersion and select the specific band, whether surface or bulk, on which superconductivity is tested.

Steady advances of ARPES apparatus have made it possible to detect superconducting gaps ranging from approx 30~meV in the cuprates a couple of decades ago down to nearly two orders of magnitude lower values nowadays \cite{Zhong2023,Fltotto2018,Wu2024,Kushnirenko2020, Lou2022, OurNature, Changdar}. The experiment is usually a ‘state of the art’ and is of course more demanding than simply pressing the button on a standard device. 

Planning the effective experiment to detect the superconducting energy gap requires an apriori knowledge of the details of the Fermi surface (FS) and underlying dispersion of a given material. Mainly because of the matrix element effects, the required portions of the FS or other k-space can only be probed with certain photon energies, polarizations and geometries. It also matters how to prepare the atomically clean surface and how good $k_z$-resolution for this termination turns out to be. Intrinsic line shape, presence of the background, overlapping features or possible anisotropy of the gap will certainly influence the result too.


The ARPES instrumentation poses a number of other serious limitations for those seeking to unravel the mysteries of superconductivity. The photons should be delivered in sufficient quantity one after the other (the continuity of the source helps to minimize space charge effects) to a specific, very small spot at the surface. They should have the same energy and ideally all of the mentioned above not varying in time. The analyzer should then be able to detect electrons originating from the same spot with sufficient energy and angular resolution, preferably in a large solid angle simultaneously. Finally, the cryomanipulator should keep the sample at a given temperature and in a certain spatial/angular position in free from external electric and magnetic fields vacuum. Neglecting any of the above factors can lead to a failure. 


If superconductivity is to be studied by ARPES in Dirac and Weyl semimetals, such as t-PtBi$_2$, the experimental routine and requirements become even more complex. This is because the system contains two distinct electronic subsystems: bulk states, which exhibit pronounced $k_z$-dispersion, and surface states. Each of these can appear very differently in ARPES spectra, requiring careful experimental design and interpretation.

Recently, another ARPES study confirmed the presence of topological Fermi arcs in t-PtBi$_2$ using a 7~eV pulsed laser \cite{KaminskiPtBi2}. However, the authors did not observe any significant shift in the leading edge under different experimental conditions. In contrast, we now routinely observe superconducting gaps on the Fermi arcs using a 6~eV continuous-wave (CW) laser \cite{OurNature, Changdar}. We initiated the present work to understand which of the factors discussed above are crucial for a successful experiment. We have collected the data using the synchrotron radiation to allow for slightly relaxed, in terms of resolution and bulk contribution, parameters and still were able to detect the gap of the order of 1-1.2~meV. Using the model of the superconducting spectral function, we show that further deterioration of experimental conditions significantly impedes the observation of the gap.

\section{Materials, methods and computational details}

Trigonal PtBi$_2$ single crystals were grown using the self - flux method. Elemental Bi (Bi pieces, 5N, chemPur) and Pt (Pt granules 1-6 mm, 4N, evochem, Advanced Materials) in a molar ratio of Pt:Bi = 1:6, were loaded in a preheated alumina crucible. The loaded crucible together with a sieve and a catch crucible were sealed under high vacuum in a quartz ampoule and placed vertically in a box furnace. The ampoule was then heated to 800 C over 8 h, held at this temperature for 36 h and then slowly cooled down to 450 C with a rate of 2 C/h. At this temperature the flux was removed using a centrifuge and the ampoule was immediately quenched in iced-water. The purity, structure and morphology of the single crystals were characterized using X-ray based techniques. Additionally, magnetic and magneto-electric measurements were performed on the samples. Our results proved to be in good agreement with previous reports.

ARPES measurements were carried out on the $1^3$ ARPES endstation  at BESSY II synchrotron (Helmholtz-Zentrum Berlin) \cite{BorisenkoOneCube}. 
Samples were cleaved \textit{in situ} with pressure better than $1 \cdot 10^{-10} $ mbar.
Experimental data was obtained using horizontally polarized synchrotron light in range from 20 to 13~eV using Scienta Omicron DA30 electron analyzer. Angular resolution was set to 0.2–0.5° and energy resolution to 2–4~meV.

\section{Results and discussion}

When attempting to observe the opening of a superconducting gap in ARPES with access to different photon energies, one usually chooses the lowest possible energy to achieve the best resolution. However, selecting the optimal photon energy for observing superconductivity in the surface states of t-PtBi$_2$ is not straightforward due to the limited $k_z$ resolution in ARPES. Although the Fermi arcs are intrinsically surface states and do not vary with photon energy (aside from possible matrix element contributions), the three-dimensional nature of the bulk features, among other factors, can still influence the accuracy of superconducting gap measurements on the Fermi arcs.
Theoretical calculations \cite{VeyratTransport, Vocaturo2024} have shown that the Fermi surface of t-PtBi$_2$ completely vanishes near the $\Gamma$-M-K plane, while near the A point $k_z$ dispersion is rather weak, which is also observed in ARPES measurements.

Figures \ref{fig1}(b,c) display Fermi surface maps measured using 20 and 17~eV photons, respectively, revealing sharp and prominent features from surface and bulk. Improved crystal quality allowed us to resolve the detailed dispersion of the Fermi arcs (see zoom-in in Fig. \ref{fig1}(k)), which is in a good agreement with theoretical calculations in Fig.\ref{fig1}(a). Meanwhile, the Fermi surface map measured at 13~eV, shown in Fig. \ref{fig1}(d), appears smeared, exhibiting broadened bulk features whose spectral intensity overlaps with the $k_z$-momentum independent Fermi arcs. This is consistent with the fact that 13~eV photons probe the electronic structure near the $\Gamma$ point while 17 and 20~eV close to the A point along the $k_z$ direction, as shown in the Fig. \ref{fig1}(m). Due to the finite $k_z$ resolution of the ARPES method and strong three-dimensionality in this region, the bulk band structure around $\Gamma$ point appears smeared, lacking any sharp, well-defined features.

Figures \ref{fig1}(e-h) and \ref{fig1} (i,j) present the dispersion of the Fermi arcs measured at 17~eV and 13~eV, respectively. Spectra measured at 17~eV clearly show that arc intersect the Fermi level only in point forming an open contour. At 13~eV, the intensity of the broad bulk states is comparable to that of the arc, whereas at 17~eV the arc is more clearly separated from the bulk. EDCs taken at the Fermi momentum $k_F$ (see Fig. \ref{fig1}(l)) show a bulk-to-arc peak intensity ratio of approximately 1:1 at 13~eV. In contrast, at 17~eV this ratio is significantly reduced thus providing a more favorable condition for superconducting measurements, where signal-to-background ratio are critical. For this reason, all subsequent temperature-dependent measurements were performed using 17~eV photons.

There are several ways to detect the opening of a superconducting gap in ARPES. These include tracking the shift of the leading edge or the peak position of the $k_F$-EDC as a function of temperature, directly observing the Bogoliubov quasiparticle dispersion, and using symmetrization. While the latter two methods provide only qualitative results, the former also yields quantitative information. This is true, of course, provided that the light source produces a sharp and stable frequency over the timescale of the experiment. Synchrotron radiation or the continuous-wave (CW) laser \cite{Felix_laser} used in our previous PtBi$_2$ report \cite{OurNature} are examples of such light sources.

In Fig. \ref{fig2}, we present temperature-dependent ARPES measurements aimed at observing the opening of the superconducting gap. Figures \ref{fig2}(a,b) show the ARPES spectra measured at 1.5~K and 15~K, respectively. The measurement direction is the same as in Fig. \ref{fig1}(g). Arrows indicate the positions of the bulk and arc EDCs, which are shown in Figures \ref{fig2}(g) and \ref{fig2}(h). The spectra were recorded over multiple cycles, three times at both low and high temperatures, to eliminate the possibility of experimental error.

As expected, the bulk state EDCs do not exhibit superconducting behavior: the leading-edge shift $\Delta_\text{LEG} $ is less than 0.1~meV, and the spectral shape resembles a Fermi function. In contrast, Fermi arc's $k_F$-EDCs show clearly $\Delta_\text{LEG} \approx 1$~meV. In the first temperature cycle, a coherence peak, characteristic of superconductivity, is also observed, though it becomes weaker in subsequent cycles — possibly due to contamination of the sample surface. In these images bulk and arc EDCs are normalized on the background. 

In Figure \ref{fig2}(j), we track the positions of the leading edge and the coherence peak extracted from the spectra shown in zoomed-in Figures \ref{fig2}(c) and \ref{fig2}(d). The data demonstrate that even under relaxed experimental conditions, it is still possible to observe signatures of the BdG dispersion, although less distinctly than with the CW laser source in our previous work. Thus, we unambiguously demonstrate the opening of the superconducting gap on the Fermi arc, while the bulk EDCs show no shift of the leading edge.

It is important to note that the leading-edge shift  does not directly reflect the size of the superconducting gap. Instead, it is a complex quantity influenced by various factors, including temperature, self-energy effects, energy resolution, background signal, and other factors. To estimate the real size of the superconducting gap, we constructed a model of the ARPES spectral function. We assumed a BCS-type superconductivity and Fermi-liquid self-energy. The model parameters were chosen to reproduce the experimental $k_F$-EDCs, as shown in Figs. \ref{fig3}(j,k). Refer to the Supplementary Information for detailed model description. 

Figures \ref{fig2}(e,f) and \ref{fig2}(i) show the modeled ARPES spectra as well as  $k_F$-EDCs for high and low temperatures. Even though the superconducting gap in the model was set to 1.2~meV, the detected $\Delta_\text{LEG}$ is only 1~meV, as in the experiment above, showing the impact of limited resolution and background signal on the leading edge shift. 

To explore the influence of resolution and background signal on superconducting gap measurements in more detail, we modeled three different cases, shown in the three columns of Fig.~\ref{fig3}. The spectral function after multiplying by the Fermi-Dirac distribution was convoluted with a two-dimensional Gaussian function with different width values $ \sigma$, corresponding to different experimental energy resolutions.

The first case corresponds to a stable 6~eV CW-laser source \cite{Felix_laser, OurNature}, delivering photons almost without background contributions and high signal-to-noise ratio. 
The second case represents a more relaxed setting in terms of resolution and includes some background, to make it similar to synchrotron ARPES experiment in this paper. 
The third column illustrates an even worse overall resolution and a background-to-signal ratio of 1:1, which is expected when using pulsed laser sources with space charge effects, discharge unfocused lamps, etc. This higher background level mimics a "bad" photon energy — one where the bulk states overlap with the arc. This situation corresponds to a region closer to the $\Gamma$ point in terms of $k_z$, as shown, for example, in our 13~eV photons measurements in Fig.\ref{fig1}(d,l).

In the bottom row of Fig.~\ref{fig3}(j,k), we demonstrate the quality of our model by comparing its $k_F$-EDCs with experimental $k_F$-EDCs under the corresponding conditions. The signal-to-background ratio and resolution were adjusted to match those observed experimentally.

As expected, even though the gap value in the model is fixed at $\Delta = 1.2$~meV across all three cases, the resulting $\Delta_\text{LEG}$ and peak positions differ significantly. In the best-resolution case, the $\Delta_\text{LEG} = 1.25$~meV closely matches the input gap value. Minor discrepancies arise due to the temperature dependence of the Fermi-liquid self-energy. The ARPES spectra shows a clear Bogoliubov quasiparticle dispersion.

Under more relaxed conditions, similar to those measured in this report, $\Delta_\text{LEG}$ becomes significantly smaller than the actual gap. The peak shift, which was observable under better conditions, becomes nearly undetectable. The Bogoliubov quasiparticle dispersion is weak but can still be extracted through fitting procedures.

Finally, in the worst-case scenario, with more moderate resolution and a nearly 1:1 background-to-signal ratio, neither the peak shift nor the leading-edge shift is visible, and no trace of Bogoliubov quasiparticle dispersion can be observed.

In conclusion, we performed synchrotron-based ARPES measurements on t-PtBi$_2$. Owing to improved sample quality, we were able to resolve the arc dispersion even at synchrotron resolutions and found good agreement with theoretical predictions. Photon energy-dependent measurements revealed that, due to strong $k_z$ dispersion, the choice of photon energy is crucial for successfully detecting the superconducting gap. Temperature-dependent measurements at 17~eV photons showed a superconducting gap opening, with typical for superconductivity coherence peak and $\Delta_{\text{LEG}} \approx 1$~meV. Fitting the ARPES spectra also showed existence of back-bending of the arc dispersion. Lastly, we make a model of superconducting ARPES spectral function, at various resolutions and signal-to-noise ratios. Although this is only a simple model that uses a simple Fermi-liquid self-energy and assumes BCS-like superconducting dispersion, which is likely not accurate for such an unconventional 2D system it still provides useful insight. The model highlights the importance of selecting the correct photon energy and achieving good overall resolution. If these conditions are not met, strong contributions from non-superconducting bulk bands can easily obscure the surface superconducting gap.

\section{Acknowledgments}
We thank the Helmholtz-Zentrum Berlin für Materialien und Energie for the allocation of synchrotron radiation beamtime. A.K. acknowledges the support of Deutsche Forschungsgemeinschaft through Project No. 555830981.
S.C. was supported by BMBF funding through project 01DK240008  (GU-QuMat).
\nocite{*}

\bibliography{apssamp}
\afterpage{
\clearpage
\begin{figure}
    \centering
    \includegraphics[width=1\linewidth]{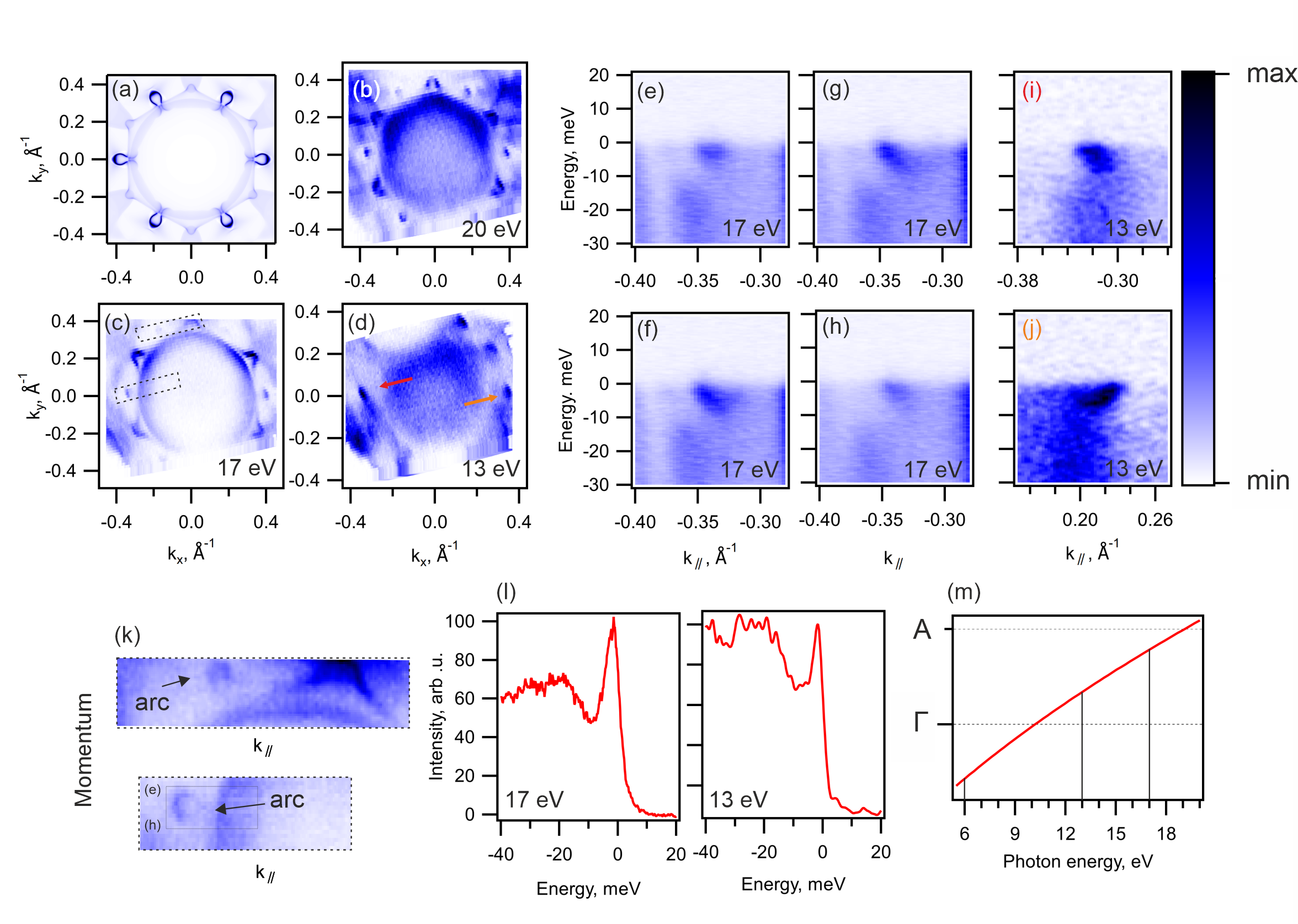}
    \caption{Fermi surface maps and energy-momentum intensity distributions for different photon energies measured at 1.5 K.  (a) Theoretical calculation of projected Fermi surface with Fermi arcs. Figure adopted from \cite{OurNature}. (b)-(d)  Fermi surface maps measured at 20, 17 and 13~eV photons respectively. (e)-(h) Energy-momentum intensity distributions in vicinity to the Fermi arc measured at 17~eV. (i)-(j) Same for the 13~eV. Direction of the measurement is shown by the red and orange arrows in (d). (k) Zoomed-in patches of the Fermi surface in (c) showing Fermi arc. (l) $k_F$-EDC from 17 and 13~eV. (m) $k_z$ to photon energy correspondence for the momentum location of the Fermi arc.}
    \label{fig1}
\end{figure}

\begin{figure}
    \centering
    \includegraphics[width=0.9\linewidth]{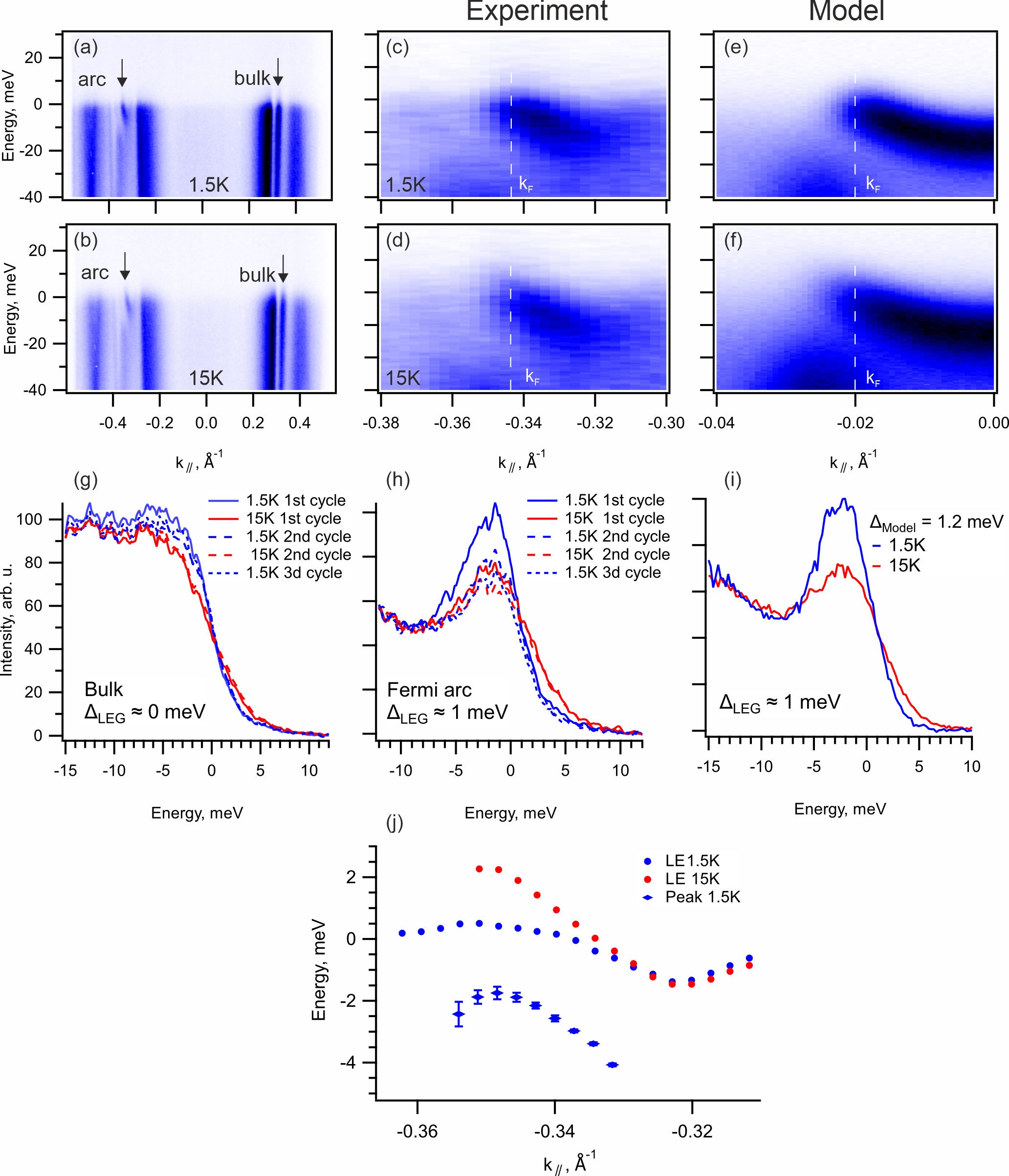}
    \caption{Temperature cycling between 1.5K and 15K. (a)-(b) Energy-momentum intensity plot at 1.5 and 15 K measured with 17~eV photons. Direction of the measurements corresponds to \ref{fig1}(g). (c)-(d) Zoomed in picture of the Fermi arc at 1.5 and 15K. (e)-(f) Model at 1.5 and 15 K. (g) EDCs taken from the bulk feature in temperature cycling. (h) $k_F$-EDCs of the arc from temperature cycling. (i) $k_F$-EDCs from the model. (j) Leading edge (LE), and peak position of the Fermi arc.}
    \label{fig2}
\end{figure}

\begin{figure}
    \centering
    \includegraphics[width=0.8\linewidth]{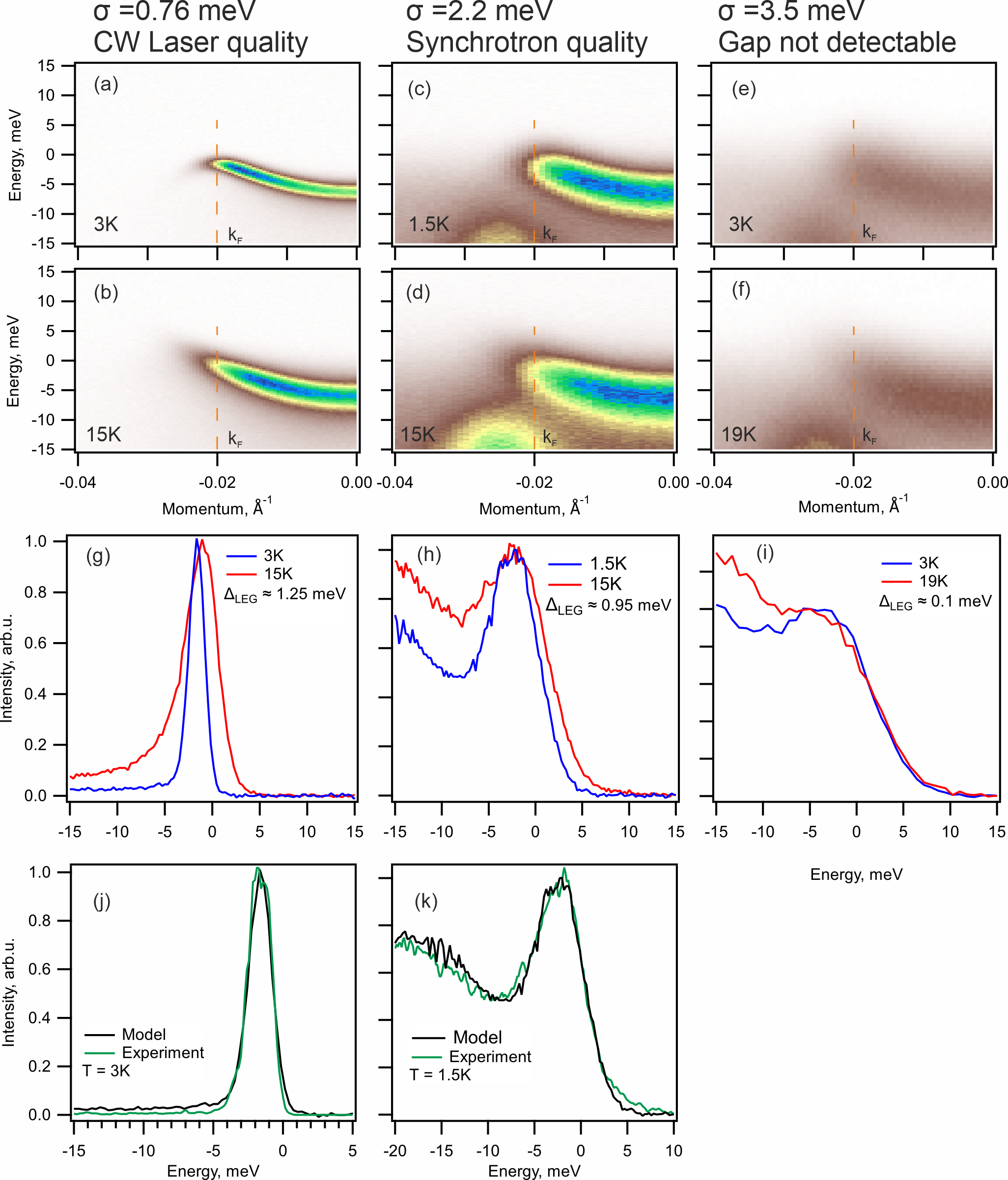}
    \caption{Model of the opening of superconducting gap ($\Delta$ = 1.2~meV) for various resolutions and signal/background ratios. (a),(b) Model for 1.5 and 15K for laser-like spectra quality. (c)-(d) Same, for synchrotron-like quality spectra. (e)-(f) Model with the parameters, where gap is already not observable. (g)-(i) $k_F$-EDCs normalized on peak maximum for corresponding models above. (j) Comparison of model above with the experimental EDC measured with 6.0~eV laser at 3K. (k) Same, but with synchrotron spectra measured at 17~eV and 1.5~K  }
    \label{fig3}
\end{figure}

\clearpage
\section{Supplementary information}
In the model, to simulate the photocurrent ($I$) measured in the ARPES experiment, we used the spectral function

\begin{equation} 
A_\text{arc}(\omega, T) \propto \frac{u_k^2\Sigma''(\omega,T)}{\left [\omega - \sqrt{\epsilon_0(k)^2 + \Delta^2} \right]^2 + \Sigma''(\omega,T)^2} + \frac{v_k^2 \Sigma''(\omega,T)}{\left[\omega + \sqrt{\epsilon_0(k)^2 + \Delta^2} \right]^2 + \Sigma''(\omega,T)^2} . 
\end{equation}

Here, $\Sigma''(\omega,T) = \alpha \omega^2 + \beta (k_BT)^2$ is the Fermi liquid imaginary part of the self-energy. The real part of the self-energy $\Sigma'$, which only represents renormalization, is omitted. Functions $v_k$ and $u_k$ are the superconducting coherence factors, and $\epsilon_0(k)$ is the bare electron dispersion, which was modeled by the quadratic function $\epsilon_0(k) = e_0(1-k^2/k_F^2)$.

To add a non-superconducting contribution from the bulk band, we used non superconducting spectral function
\begin{equation} 
A_\text{bulk}(\omega, T) \propto \frac{\Sigma''(\omega,T)}{\left[\omega - \epsilon_1(k) \right]^2 + \Sigma''(\omega,T)^2} , 
\end{equation}
with bare electron dispersion $\epsilon_1(k) = e_1(1+k^2/k_{1}^2)$. The total spectral function is the sum of the superconducting arc and the non-superconducting bulk contributions $A_\text{total} = A_\text{arc} + A_\text{bulk}$.

Further, the spectral function $A_\text{total}$ is multiplied by the Fermi-Dirac distribution $f(\omega, T)$ and convoluted with a 2D Gaussian function $R_{k,\omega} \propto \exp \left[- \frac{\omega^2}{2 \sigma^2} - \frac{k^2}{2 \sigma_k^2}\right]$ to simulate experimental energy and momentum resolution. Lastly, we add a momentum-independent background using the empirical formula $B(\omega, T) = f(\omega + \omega_b, T+T_b)$.

The total simulated photocurrent has the form

\begin{equation}
    I(\omega,T,\Delta) = \left\{ \left[M_1A_\text{arc}(\omega,T,\Delta)  + M_2A_\text{bulk}(\omega,T)\right]f(\omega,T) \right\}\otimes R_{k,\omega} + B(\omega, T).
\end{equation}
Coefficients $M_1$ and $M_2$ adjust signal-to-background ratio and are set to replicate the experimental curves. 

The parameters of the model are the following:
$\alpha = 0.02$~meV$^{-2}$, $\beta = 0.05$~meV$^{-2}$, $\Delta = 1.2$~meV, $e_0 = 6$ meV, $k_F = -0.02$~\AA, $\omega_b = 3$~meV, $T_b = 22$~K, all the same across three cases. 
}

\end{document}